\documentclass[aps,prl,twocolumn,superscriptaddress]{revtex4-1}
\pdfoutput=1
\usepackage{graphicx}
\usepackage{amsmath,amsfonts,amssymb}
\usepackage[colorlinks=true,linkcolor=blue,citecolor=blue]{hyperref}

\begin{document}

\title{Point contacts in encapsulated graphene}


\author{Clevin Handschin}
\affiliation{Department of Physics, University of Basel, Klingelbergstrasse 82, CH-4056 Basel, Switzerland}
\affiliation{Swiss Nanoscience Instute, Klingelbergstrasse 82, CH-4056 Basel, Switzerland}

\author{Balint F\"{u}l\"{o}p}
\affiliation{Department of Physics, Budapest University of Technology and Economics and Condensed Matter Research Group of the Hungarian Academy of Sciences, Budafoki ut 8, 1111 Budapest, Hungary}

\author{P\'eter Makk}
\affiliation{Department of Physics, University of Basel, Klingelbergstrasse 82, CH-4056 Basel, Switzerland}

\author{Sofya Blanter}
\affiliation{Department of Physics, University of Basel, Klingelbergstrasse 82, CH-4056 Basel, Switzerland}

\author{Markus Weiss}
\affiliation{Department of Physics, University of Basel, Klingelbergstrasse 82, CH-4056 Basel, Switzerland}

\author{K. Watanabe}
\affiliation{National Institute for Material Science, 1-1 Namiki, Tsukuba, 305-0044, Japan\\}

\author{T. Taniguchi}
\affiliation{National Institute for Material Science, 1-1 Namiki, Tsukuba, 305-0044, Japan\\}

\author{Szabolcs Csonka}
\affiliation{Department of Physics, Budapest University of Technology and Economics and Condensed Matter Research Group of the Hungarian Academy of Sciences, Budafoki ut 8, 1111 Budapest, Hungary}

\author{Christian Sch\"onenberger}
\email{Christian.Schoenenberger@unibas.ch}
\affiliation{Department of Physics, University of Basel, Klingelbergstrasse 82, CH-4056 Basel, Switzerland}

\date{\today}

\begin{abstract}
We present a novel method to establish inner point contacts on hexagonal boron nitride (hBN) encapsulated graphene heterostructures with dimensions as small as 100 nm by pre-patterning the top-hBN in a separate step prior to dry-stacking. 2 and 4-terminal field effect measurements between different lead combinations are in qualitative agreement with an electrostatic model assuming point-like contacts. The measured contact resistances are 0.5-1.5 k$\Omega$ per contact, which is quite low for such small contacts. By applying a perpendicular magnetic fields, an insulating behaviour in the quantum Hall regime was observed, as expected for inner contacts. The fabricated contacts are compatible with high mobility graphene structures and open up the field for the realization of several electron optical proposals.
\end{abstract}


\maketitle


In recent years several experiments have shown, that ballistic graphene is an ideal platform for electron optical experiments. These experiments included the observation of Fabry Perot resonances \cite{Rickhaus13,Grushina13, Campos2012}, snake states \cite{Rickhaus2015,Taychatanapat2015}, electron guiding \cite{guiding15}, magnetic focusing \cite{Taychatanapat2013,Calado_APL} or ballistic Josephson currents \cite{Mizuno13,Calado15,Shalom15, Allen15}.

Up to now graphene encapsulated in hBN, which yields the highest quality for graphene on substrate, could only be accessed by top- or side-contacts from the edge of the device. \cite{Kretinin14, Wang13} However, a different type of contacts, inner point contacts (PCs) are required in order to realize  several theoretical proposals on graphene such as e.g. the Veselago lensing \cite{Veselago68} in single \citep{Cheianov07, Gomez12, Veselago15} layer graphene, valley \cite{Vesperinas08, Peterfalvi12} or spin focussing \cite{Zareyan10} in graphene or for the investigation of skipping and snake orbits of charge carriers at a pn-junction in combination with a strong magnetic field perpendicular to the graphene plane \cite{Patel12, Davies12}.

In order to make PCs in the middle of the graphene sheet evaporated, sputtered or atomic layer deposited (ALD) dielectrics, such as MgO, SiO$_2$ or Al$_2$O$_3$ have been used so far \cite{Zaho12}. However, these materials are inferior to the layered material hBN when it comes to the preservation of the graphene quality \cite{Dean10}. It is possible to establish PCs on graphene using a STM tip where also the position of the contact is changeable. However, doing this at low temperatures, involving several PCs at the same time is extremely challenging.

Here we present a novel method to establish PCs to graphene encapsulated in hBN by pre-patterning the h-BN flake before the stacking process. This allows to access the graphene at arbitrary position with contacts smaller than 100 nm in diameter. The method is compatible with clean graphene fabrication, since the graphene transport channel will not be in contact with any resists or solvents during fabrication \cite{Wang13,Zomer14,Gomez14, Banszeruse1500222rohtua}. We extract the graphene quality by comparing measured 2 and 4 terminal resistance values with a simple model. Furthermore we show localization of the edge-states around the PCs in a magnetic field, expected for a proper inner contact in the quantum Hall regime.

\begin{figure}[htbp!]
    \centering
      \includegraphics[width=1\columnwidth]{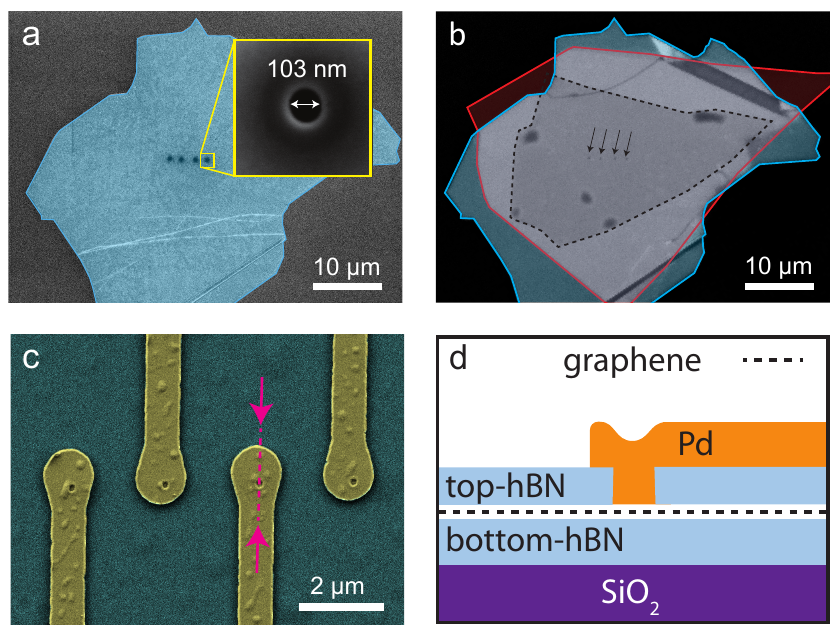}
    \caption{\textbf{Fabrication of hBN-graphene-hBN hetero-structures with top-PCs. a,} False color SEM image of the top-hBN on SiO$_2$ substrate after drilling the holes with the Ga-FIB. Inset: Close-up of a single hole having a diameter of $\sim$100 nm. \textbf{b,} Optical image of a final stack. The different layers are indicated in red (bottom hBN), black (graphene) and blue (top hBN). The holes are indicated with arrows.  \textbf{c,} False color SEM image of the final stack (blue) with Pd contacts (yellow) overlapping the drilled holes. \textbf{d,} Schematic of the cross-section as indicated in (\textbf{c}) with pink arrows and the dashed line.}
    \label{fig:figure_fabrication}
\end{figure}


To produce PCs compatible with high mobility graphene we have fabricated four equally spaced holes in the top hBN layer of hBN-graphene-hBN hetero-structures. In order not to damage the graphene, the top-hBN layer ($\sim$10-20 nm thick) was pre-patterned on a separate SiO$_2$ wafer using a gallium based focused ion beam (Ga-FIB, see details in supporting) prior to the dry-stacking of the heterostructure \citep{Wang13, wangWSe2}. In contrast to establishing the holes with conventional e-beam lithography and subsequent etching, the drilled holes are better defined in shape and the diameter can be adjusted more reliably. In our samples we use a hole-diameter of approximately 100 nm as it is shown in Fig. \ref{fig:figure_fabrication}a. Before picking-up the top-hBN from the SiO$_2$ wafer, it is briefly exposed to a CHF$_3$/O$_2$ plasma treatment as without the plasma the flakes are pinned to the SiO$_2$ and pick-up is not possible.

The top-hBN flake is transferred from the SiO$_2$ substrate to a 1 $\mu$m thick poly-propylene-carbonate (PPC) polymer by spinning the PPC directly onto the SiO$_2$ chip containing the top-hBN with the holes  and then peeling the polymer off. The remaining assembly procedure of the hBN-graphene-hBN heterostructure follows the dry-stacking approach proposed by L. Wang \textit{et al.} \citep{Wang13}. Since  only the top side of hBN comes in contact with the polymer, the method preserves the clean fabrication of dry-stacking graphene. The bottom hBN flake is exfoliated to the target wafer directly, with typical thickness of 20-30$\,$nm.

We use a strongly doped Si$^{++}$/SiO$_2$ wafer with a 300 nm thick oxide to gate our devices. The final stack was annealed in forming gas at $T$=300 $^\circ$C for 3 hours in order to reduce strain and maximize the areas without bubbles \cite{Haigh2012}. A contrast adjusted optical image of the annealed stack is shown in Fig.~\ref{fig:figure_fabrication}b. The 100 nm thick palladium (Pd) contacts are established using standard e-beam lithography and e-gun evaporation. A false color SEM image of the contact area is shown in Fig.~\ref{fig:figure_fabrication}c. A schematic of the cross-section of the stack with contacts, as indicated with the dashed pink line in Fig.~\ref{fig:figure_fabrication}c, is shown in Fig.~\ref{fig:figure_fabrication}d. Further details of the fabrication are given in the Supporting Material.\\
In total 4 different samples were produced all showing a similar behaviour. The measurements were performed at cryogenic temperatures (1.5-4 K) using standard low-frequency lock-in technique.

In the following, first the contact resistance and the field-effect measurements at zero magnetic field are discussed. In the second part, the behaviour of the devices at high magnetic field perpendicular to the graphene plane is presented. For the calculation of the charge carrier mobility, a model fitting the device geometry is introduced. The nomenclature of the various differential resistances is given as $R_{ij,nm}=dV_{nm}/dI_{ij}$, with $I_{ij}$ flows from PC $i \rightarrow j$ and the voltage $V_{nm}$ measured as the  difference between PC $n$ and $m$, with $i,j,n,m \in 1-4$.

\begin{figure}[htbp]
    \centering
      \includegraphics[width=1\columnwidth]{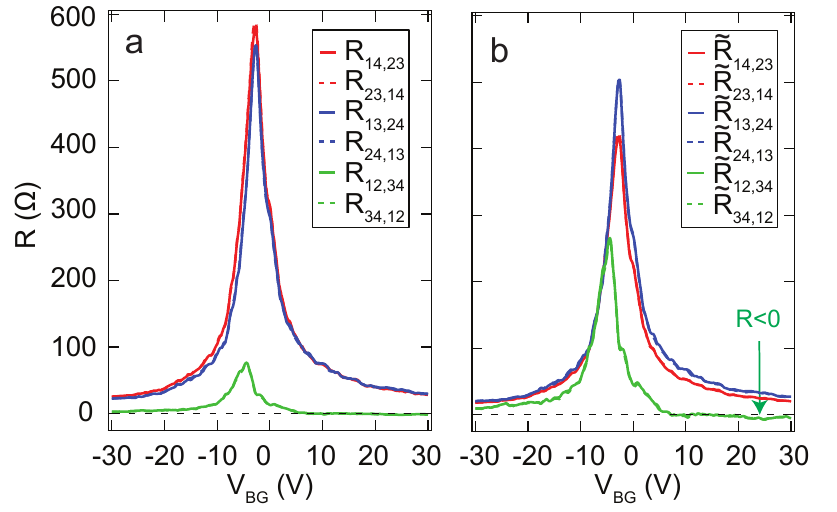}
    \caption{\textbf{4-terminal measurement of the graphene resistance between different combinations of PCs. a,} The original 4-terminal resistance $R_{ij,nm}$ measured between the six different lead combinations. Measurements with inverse voltage and current probes are identical. \textbf{b,} Resistance of the same data presented in (\textbf{a}) multiplied by a factor depending on the measurement geometry  as given by Eq.~\ref{eq:7}. In the non-local measurement $R_{12,34}$, the voltage $V=V_{n}-V_{m}$ was negative in the n-doped region.}
    \label{fig:figure_measurements1_1}
\end{figure}

Figure~\ref{fig:figure_measurements1_1}a shows $R_{ij,nm}$ for all possible 4-terminal configurations. Out of the six possible configurations, only three are independent: measurements where current- and voltage-probes are inverted are identical as expected from the Onsager relations \cite{Onsager31}. The resistance traces show a sharp maximum around zero doping, corresponding to the charge neutrality point (CNP) of graphene.

\begin{figure}[htbp]
    \centering
      \includegraphics[width=1\columnwidth]{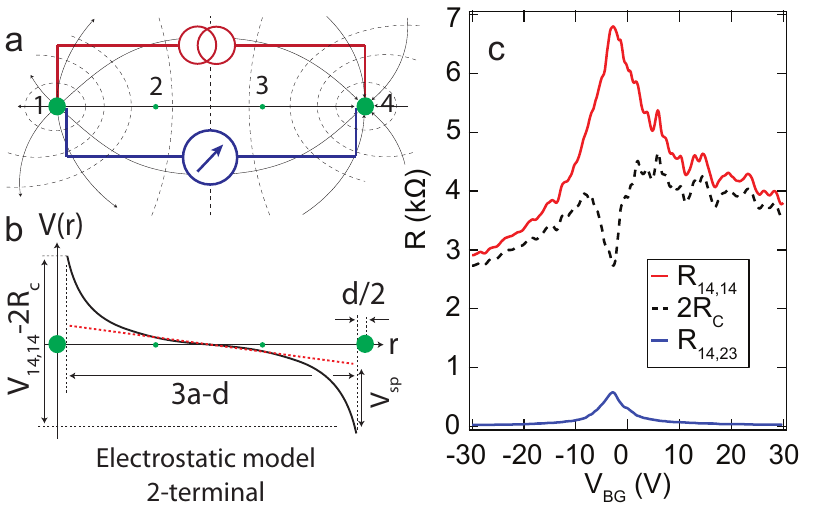}
    \caption{\textbf{Contact resistance extracted from 2- and 4-terminal measurements. a,}  The 2-terminal measurement setup and a sketch of the corresponding electric field profile. The 2-terminal measurement is performed between the two outer-most PCs which have a distance of $3a$ from each other, whereas the reference four terminal measurement between contacts 2 and 3. \textbf{b,} The red, dashed line represents the linear interpolation of $V_{23}$ compared to the actual electrostatic potential (black, solid line) along the line connecting the contacts. The resulting potential difference at distance d/2 from source- and drain-center is indicated with $V_{sp}$. \textbf{c,} 2-terminal resistance between 1 and 4 is shown in red, whereas the 4-terminal resistance $R_{14,23}$ is shown in blue. Extracted contact resistance of the configuration given in (\textbf{a}) using Eq.~\ref{eq:10} and \ref{eq:13}}.
    \label{fig:figure_measurements1_2}
\end{figure}

For rectangular graphene devices, where the current density within the graphene sheet is constant, the mobility $\mu$ can be deduced by measuring the field effect of the longitudinal resistance $R_{xx}(V_{BG})$, taking the length and width of the Hall bar into account. For PCs, which are situated in the middle of the graphene sheet, a different formula has to be used since the current density within the graphene sheet varies.
Here we introduce a model to extract the sheet conductivity ($\sigma$) from the 4-terminal measurement of the resistance, assuming an infinite graphene sheet with a constant $\sigma$. The four contacts are at positions $\mathbf{{r_{x}}}$ with $x=i,j,n,m$ and diffusive transport.
Starting from a single PC at position $\mathbf{r_{i}}$, the current spreads isotropically into the graphene. This leads to a current density of $\mathbf{j(r)}=I/(2\pi |\mathbf{r}-\mathbf{r_{i}}|)$ at distance $\mathbf{|r-{r_{i}}|}$ away from the PC, where $I$ is the current. According to $\mathbf{j(r)}=\mathbf{E(r)}\sigma$, the electric field, $\mathbf{E(r)}\sim 1/|\mathbf{r}-\mathbf{r_{i}}|$, leading to an electrostatic potential $V(\mathbf{r})\sim \ln\left(|\mathbf{r}-\mathbf{r_{i}}|\right)$. Assuming a current flow between two PCs from $i\rightarrow j$, the potential at position $\mathbf{r}$ is obtained by the superposition principle:
\begin{equation} \label{eq:1}
V(\mathbf{r})= -\frac{I_{ij}}{2\pi\sigma}*\ln\left(|\mathbf{r}-\mathbf{r_i}|\right)+\frac{I_{ij}}{2\pi\sigma}*\ln\left(|\mathbf{r}-\mathbf{r_j}|\right)+C,
\end{equation}
where C is an integration constant. In the 4-terminal measurement only the voltage difference between the two leads at position $n$ and $m$ ($V_{nm}=V(\mathbf{r_{n}})-V(\mathbf{r_{m}})$) is measured. For simplicity we assume that the voltage probes do not influence the electric field pattern in graphene as shown in Fig.~\ref{fig:figure_measurements1_2}a and b. This leads to
\begin{equation} \label{eq:7}
V_{nm}=\frac{I_{ij}}{2\pi\sigma}*\ln\left(\frac{|\mathbf{r_n}-\mathbf{r_j}|}{|\mathbf{r_n}-\mathbf{r_i}|}\frac{|\mathbf{r_m}-\mathbf{r_i}|}{|\mathbf{r_m}-\mathbf{r_j}|}\right),
\end{equation}
where the conductivity $\sigma$ can now be extracted from the measurement.

Using $R_{ij,nm}$ $\sigma$ can be deduced. The mobility, $\mu$ of the graphene was then extracted using the linear dependence of $\sigma(n)$ on the carrier density, $n$ ($\sigma=ne\mu$). The density was calculated from the back gate voltage using a parallel plate capacitor model. The hole and electron doped region revealed mobilities of $\mu_{h}\sim$35'000\,cm$^2$/(Vs)  and $\mu_{e}\sim$25'000\,cm$^2$/(Vs) respectively. This is in good agreement with the conversion extracted from the evolution of the filling factors in a quantum Hall experiment (QHE) (see supplementary information). An alternative way to determine the mobility is by the onset of the Shubnikov de Haas (SdH) oscillations in the QHE measurement. According to $\mu B=\omega_{c} \tau=1$, where $\omega_{c}$ is the cyclotron frequency and $\tau$ is the scattering time, the charge carriers can complete a full cyclotron orbit before being scattered. With the SdH oscillations starting at $\sim$0.4-0.5$\,$T, a corresponding mobility of $\mu \sim$20'000-25'000\,cm$^2$/(Vs) is extracted which is in good agreement with the values deduced from the field effect measurements.
Confirmation that a single layer graphene (SLG) is encapsulated in hBN was given by the observed sequence of  filling factors in magnetic field (see supplementary information) and by Raman spectroscopy (see supplementary information).\\

From Eq.~\ref{eq:7} it follows that in case of a homogeneous $\sigma$, all 4-terminal resistance measurements can be renormalized according to

\begin{equation} \label{eq:9}
\widetilde{R}_{ij,nm}=R_{ij,nm} \ln\left(\frac{|\mathbf{r_n}-\mathbf{r_j}|}{|\mathbf{r_n}-\mathbf{r_i}|}\frac{|\mathbf{r_m}-\mathbf{r_i}|}{|\mathbf{r_m}-\mathbf{r_j}|}\right)^{-1}.
\end{equation}

With all four contacts at equidistant spacing the logarithm in Eq.~\ref{eq:9} simplifies to ln(4), ln(3) or ln(3/4) depending on the measurement configuration. If the model with all the assumptions is valid, all $\widetilde{R}_{ij,nm}$ should be equal and given by $1/2\pi\sigma$. The difference between the original and renormalized values can be seen in Fig.~\ref{fig:figure_measurements1_1}a and \ref{fig:figure_measurements1_1}b, respectively.\\

The renormalized values $\widetilde{R}_{ij,nm}$ in Fig.~\ref{fig:figure_measurements1_1}b are not exactly overlapping, as would be expected for a perfect system given in Eq.~\ref{eq:9}. However, one can see that the non-local measurements (voltage probes outside the current path), shown in green, which were in the original data smaller by a factor of 7.5 (8) from the blue (red) curve, deviates now only by a factor of 2 (1.6) after renormalization. On the other hand, the local measurements, the blue and red curves are in rather good agreement before and after renormalization. Overall, the rescaled are much closer to each other than the unscaled ones, which confirms that our theoretical model is realistic.

The deviations from the ideal case can be related to the boundary conditions assumed for the model. The most significant deviations from the ideal model are probably i) the finite dimensions of the metallic PCs and ii) the finite size of the graphene sheet, which both change the electric field pattern. Besides that, the sheet conductivity does not seem to be fully uniform within the sample as can be seen by the slight shift of the charge neutrality point between several measurements. The charge neutrality points are at $V$=-2.6$\,$V, -2.8$\,$V and -4.8$\,$V, respectively. Moreover, a non-uniformity of the doping profile arises also from the screening of the top contacts. This results in different lever arms of the back-gate for regions covered and not-covered by the electrodes. Finally, Eq.~\ref{eq:9} was derived assuming a completely diffusive sample. However, the charge carriers in the sample are most likely in an intermediate regime between the diffusive and the ballistic regime.
Using the Drude formula, the scattering mean free path can reach 1$\,\mu$m at $V_{BG}=30\,$V, which is in the same order as the contact distance $a=2.2\,\mu$m. In this intermediate regime it can occur that the voltage drop over a larger, but clean (ballistic) area is lower compared to a smaller, but dirty (diffusive) area. Moreover, for ballistic trajectories the probability of arriving at a contact, which is farther away can be higher. This picture explains the negative resistance at certain doping values observed in the non-local measurement indicated with an arrow in Fig.~\ref{fig:figure_measurements1_1}c. For all the configurations where the voltage-probes are (at least partially) within the current path, the bias voltage will be dominant and consequently no negative signal can be seen.\\


In order to extract the contact resistance $R_c$ which arises between the metal leads and the graphene, $R_{C}$ (interface resistance), we turn to two terminal measurements. To calculate the contact resistance we measure the 2-terminal resistance $R_{2T}$ between the outer contacts (1,4) as sketched in Fig.~\ref{fig:figure_measurements1_2}a. Then the contact resistance can be calculated according to:
\begin{equation} \label{eq:10}
R_{2T}=2 R_{C}+ A \cdot R_{4T},
\end{equation}
where $A \cdot R_{4T}=A \cdot  R_{14,23}$ is the intrinsic graphene resistance including a geometry factor $A$ (which will be evaluated in the following) and $R_{2T}=R_{14,14}$. Here we assumed the same contact resistance for the two contacts. Due to the higher electric field near the source and drain contacts, shown in Fig.~\ref{fig:figure_measurements1_2}a, the voltage changes faster near them. This can be seen in Fig.~\ref{fig:figure_measurements1_2}b, where the potential along the line connecting the contacts is plotted.
The resistance coming from the non-linearity of the potential near the contacts is called spreading resistance and leads to a potential difference that is marked as $V_{sp}$ in Fig.~\ref{fig:figure_measurements1_2}b \cite{Zhang12}. It is of the same origin as Maxwell's resistance that occurs in metallic point contacts \cite{Maxwellbook}.\\

To calculate the geometrical factor $A$, we use Eq.~\ref{eq:7}, and apply it together with Eq.~\ref{eq:10} to the situation sketched in Fig.~\ref{fig:figure_measurements1_2}a. Because the potential is singular at position $\mathbf{r_{i}}$ and $\mathbf{r_{j}}$ we introduce a cut-off for the potential at $d/2$ away from the singularity, where $d$ is the diameter of the contact. Physically speaking this accounts for the equipotential within the metallic PCs with a finite dimension. Doing so, both terms in the numerator and denominator of Eq.~\ref{eq:7} become $(3a-d/2)$ and $d/2$ respectively,  where $a$ is the distance in between two neighbouring contacts. This leads to
\begin{equation} \label{eq:11}
R_{14,14}=\frac{1}{2\pi \sigma}\ln\left(\left(\frac{6a}{d}-2\right)^2\right)\; +2R_{C}.
\end{equation}
Using $R_{4T}=R_{14,23}$ where  $R_{14,23}=ln(4)/(2\pi\sigma)$ the geometrical factor becomes:
\begin{equation} \label{eq:13}
A=\ln\left(\left(\frac{6a}{d}-2\right)^2\right)\frac{1}{ln\left(4\right)}.
\end{equation}
In Fig.~\ref{fig:figure_measurements1_2}c the extracted contact resistance is shown using a geometry factor of $A$=7.04 ($a$=2.2 $\mu$m and $d$=100 nm).  The contact resistance of different contact configurations and different devices is of the order of R$_{c}$=0.5-1.5 k$\Omega$ at high doping. This value is quite remarkable for PCs of only 100 nm in diameter since as well top-contacts with significantly larger areas (in the order of $\mu$m$^2$) have resistances in the k$\Omega$ range.
Moreover, the presented model shows that the contact resistance at high doping is given roughly by the 2-terminal resistance  at high doping ($V_{BG}$=$\pm$30 V). In this case graphene becomes very conductive and the voltage-drop over the graphene is minimal, therefore $R_{2T}\sim 2 R_{C}$.

\begin{figure}[htbp]
    \centering
      \includegraphics[width=1\columnwidth]{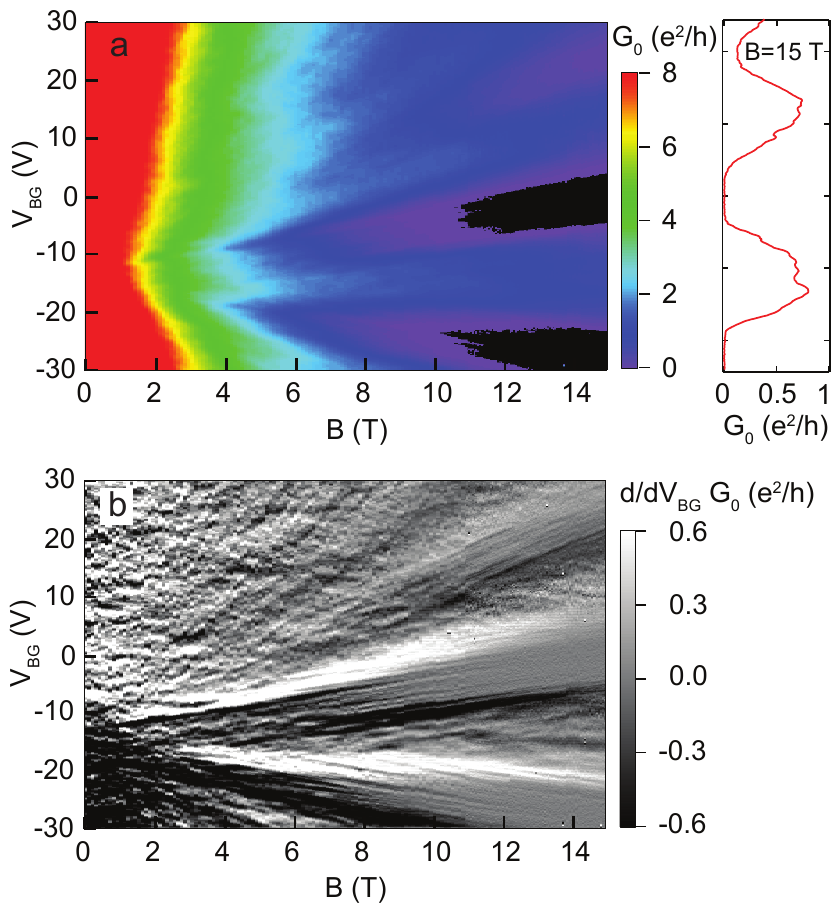}
    \caption{\textbf{2-terminal conductance  as a function of back-gate and magnetic field. a} The conductance between neighbouring PCs (distance of 1$\,\mu$m) becomes zero at high magnetic fields. The black area shows the threshold for $G<0.025$ e$^2$/h ($R>$1 M$\Omega$). A line-trace at $B=15\; T$ is given on the right-hand side of the color-plot. \textbf{b,} Numerical derivative calculated with respect to $V_{BG}$ of the measurement given in (\textbf{a}) reveals the constant-conductance regions more pronouncedly.}
    \label{fig:figure_measurements2}
\end{figure}

To further characterize the PCs, we have applied a magnetic field perpendicular to the graphene sheet, which forces the charge carriers to move along cyclotron orbits. With a sufficiently high magnetic field the device is driven into the quantum Hall regime, into a state, where the bulk of the sample is insulating, since charge carriers will be localized either around the PCs or along the edges of the sample, which decouples the PCs from each other and the edge of the sample. In the case of a homogenously doped and gated device one would expect complete insulation as soon as the cyclotron orbit and the magnetic length are smaller than the distance between the PCs. The latter is the case for magnetic fields in the few hundred mT range.
Moreover $\omega_c \tau > 1$ is required.

The conductance  as a function of back-gate and magnetic field of a sample with a hole to hole spacing of $a$=1 $\mu$m and a mobility of $\mu=$15'000 cm$^2$/(Vs) is shown in Fig.~\ref{fig:figure_measurements2}. The magnetic field dependence of the device discussed before can be found in the Supporting Information.  The black region in Fig.~\ref{fig:figure_measurements2}a shows values of $G_0<0.025$ e$^2$/h ($R>$1 M$\Omega$). It can be seen, that by the application of magnetic field the sample becomes insulating, however, the fields required are much higher than expected. From simple considerations above this should happen around $0.5-1\,$T.

Inhomogenous doping distribution in the sample can result in coupling of the contacts. For complete insulation of the device, the graphene has to be simultaneously insulating in the whole region between the contacts, since on the border of regions with different filling factors edge currents will flow. Tunneling to these edge states can give short-cut currents between the contacts.
As already mentioned in the text, doping inhomogenities exist within our sample, as can be seen from the shift of the charge neutrality point between different 2T measurements. Furthermore, locally the back-gate can be screened by the top contacts. However, our estimates have shown, that the screening changes the gate efficiency by less than 4\% far away from the CNP, where the quantum capacitance is high. Only close to the CNP, where the quantum  capacitance is small, does screening of the contacts increase to 20\%. Furthermore, an additional offset potential may emerge in the regions of the top contacts due the formation of a contact potential between the palladium contacts and h-BN. We emphasize, that substantial part of the voltage drops in the region close to the contact.
The combination of all these effects can cause local differences in the filling factor, which can account for the observed high threshold fields.

In future devices insulation of at lower field can be achieved by choosing devices with smaller doping inhomogenities, which can come from bubbles present in the stacks. Moreover the inhomogenous screening of the top contacts or offset potentials can be circumvented by careful design, in which the flake would be fully covered with a metallic plane to achieve a homogeneous doping situation.\\

We have shown a new method to establish inner point contacts with dimensions of 100 nm in a hBN-graphene-hBN heterostructure. A simple model has been introduced which qualitatively explains our 2 and 4-terminal gate dependent conductance measurements. Surprisingly low contact resistance R$_{c}$=0.5-1.5 k$\Omega$ have been found despite the small PC size. Magnetic field measurements showed that the inner contacts are decoupled from the edge and from each other at high magnetic fields.

The presented technique is compatible with high-quality encapsulated graphene, since the hBN flake is patterned prior to the stacking and therefore the graphene remains clean. With further optimization one can expect devices with mobilities around 100 000 cm$^2$/(Vs). The technique also holds the potential to further decrease the contact size, since with the Ga-FIB hole diameters below $d$=20 nm are possible.

The point contacts introduced here give the possibility to complement side and top contacts in complex devices and could be the potential milestone towards realizing novel concepts like lensing or measurements of caustics in p-n junctions.\\

\textbf{Acknowledgments}

This work was further funded by the Swiss National Science Foundation, the Swiss Nanoscience Institute, the Swiss NCCR QSIT, the ERC Advanced Investigator Grant QUEST, the ERC grant 258789, OTKA grant K112918 and the EU flagship project graphene.

The authors thank Simon Zihlmann, Srijit Goswami and Ming-Hao Liu for the fruitful discussions.


\bibliographystyle{unsrt}

\clearpage
\onecolumngrid
\section{Supplementary}

        \setcounter{figure}{0}
        \renewcommand{\thefigure}{S\arabic{figure}}%

 \subsection{Fabrication}
\begin{figure}[htbp]
    \centering
      \includegraphics[width=0.7\columnwidth]{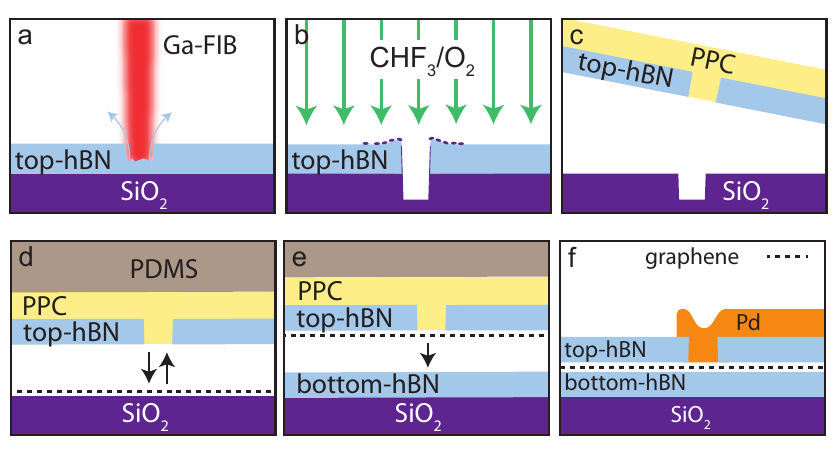}
    \caption{\textbf{Fabrication of the point contacts on a hBN-graphene-hBN heterostructure (not to scale). a,} Drilling the holes into the top-hBN using a Ga-FIB.  \textbf{b,} Due to the drilling, SiO$_{2}$, hBN and Ga is deposited on the side of the hole and in the near proximity of the hole on top of the hBN (indicated with purple dots). Exposing the top-hBN flake to a CHF$_{3}$/O$_{2}$ plasma avoids pinning of flake to the SiO$_{2}$ substrate. \textbf{c,} The top-hBN flake with the holes is removed from the SiO$_{2}$ support by spin-coating PPC on top of the wafer and then peeling it gently of.  The remaining assembly of the hBN-graphene-hBN stack is shown in the figures (\textbf{d})- (\textbf{e}) where the black arrow indicate the pick-up of graphene and the release of the half-stack on a bottom hBN respectively. After complete assembly of the stack, the PPC is dissolved in Chloroform. \textbf{f,} The palladium contacts are assembled using standard e-beam lithography and e-gun evaporation.}
    \label{fig:Supplementary_Fabrication}
\end{figure}
In order to drill holes into the top-hBN with a gallium based focused ion beam (Ga-FIB) we use an acceleration voltage of 30 keV and the smallest possible current (1.1 pA) in order to obtain highest resolution. The hBN to be patterned was exfoliated on a Si$^{++}$/SiO$_2$ substrate with a 315 nm thick oxide, using the scotch-tape technique. The chips were previously carefully cleaned using Piranha solution (98\% H$_2$SO$_4$ and 30\% H$_2$O$_2$ in a ratio of 3:1) since it is the bottom face of the hBN which will later on contact the graphene.
Once an ideal hBN flake (thickness $\sim$10-30 nm) is identified by optical microscopy, the Ga-FIB is  used to drill several  holes into the flake (with diameter $d\sim$100 nm and a equidistant spacing of 1-2.2 $\mu$m) as sketched in fig \ref{fig:Supplementary_Fabrication}a.\\
Before picking-up the hBN from the SiO$_2$ wafer, it is briefly exposed to a CHF$_3$/O$_2$ plasma (40 sccm/4sccm, 60 mTorr, 60 W, 15 s) as shown in fig. \ref{fig:Supplementary_Fabrication}b. It turned out that without exposing the hBN flakes to the plasma, the hBN flakes could not be picked-up from the SiO$_2$ substrate. A possible explanation might be that during the drilling of the holes with the Ga-FIB, SiO$_2$ from the wafer is sputtered on the side of the holes which pins the flake to the wafer. The CHF$_3$/O$_2$ plasma removes this layer and allows therefore a successful pick-up of the flake from the SiO$_2$ chip. \\
To pick-up the top-hBN, the SiO$_2$ chip is spin-coated with $\sim$1 $\mu$m of poly-propylene-carbonate (PPC) and baked at 80 $^\circ$C for 5 minutes. By peeling-off the PPC gently from the substrate (fig. \ref{fig:Supplementary_Fabrication}c), all hBN flakes are transferred from the SiO$_2$ onto the PPC polymer. Peeling-off the PPC without breaking the hBN flakes works best when slowly releasing the PPC at a low angle from the SiO$_2$ chip (drilled flakes are more likely to break).  The PPC with the hBN flake is then placed on a home-made stamp of $\sim$0.5 mm PDMS.\\
The remaining assembly procedure of the hBN-graphene-hBN heterostructure follows the dry-stacking approach proposed by L. Wang \textit{et al.} as shown in fig. \ref{fig:Supplementary_Fabrication}d-e. \citep{Wang13}.  The 100 nm thick palladium (Pd) contacts are established using standard e-beam lithography and e-gun evaporation. A cross-sectional sketch of the final device is shown in fig. \ref{fig:Supplementary_Fabrication}f.
\newpage

\subsection{Low field measurements}
\begin{figure}[htbp]
    \centering
      \includegraphics[width=1\columnwidth]{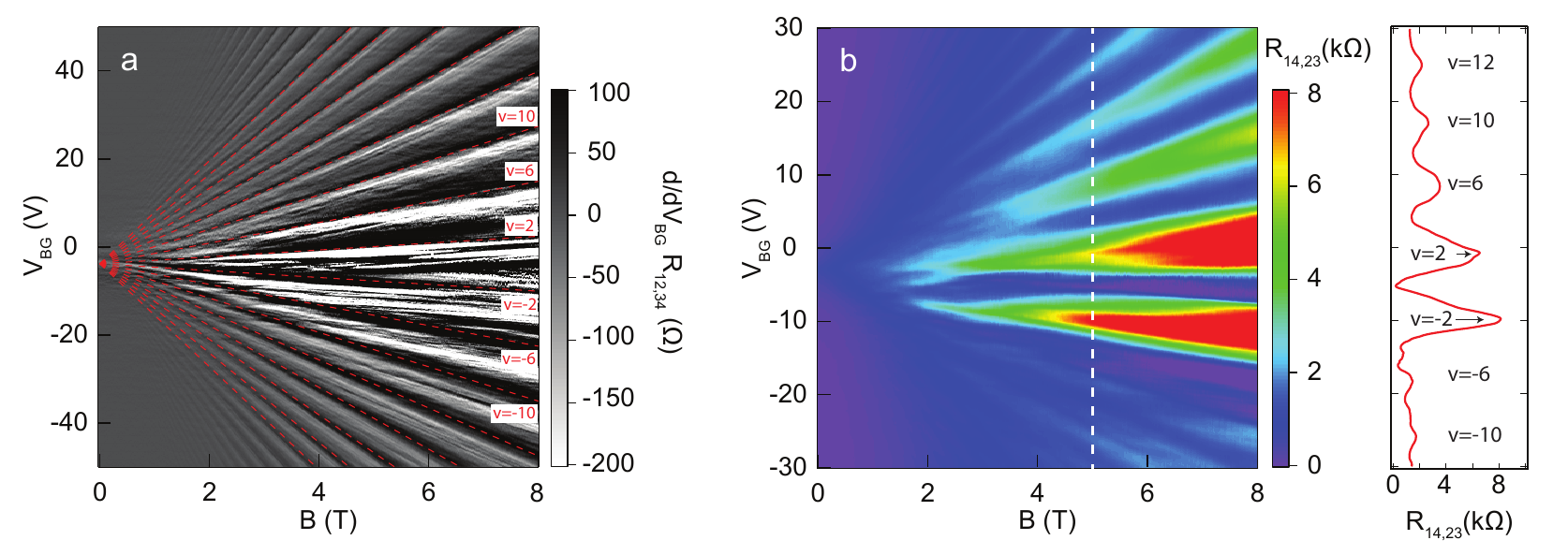}
    \caption{\textbf{4-terminal measurements as a function of back-gate and magnetic field. a,} Derivative $d/dV_{BG}\; R_{12,34}$ of a non-local measurement. The sequence of the filling-factors (red, dashed lines), $\nu=\pm$ 2, $\pm$ 6, $\pm$ 10... is in agreement with the ones expected for SLG. \textbf{b,} 4-terminal measurement of the resistance $R_{14,23}$ which is more comparable to a classical QHE. A line-cut at $B$=8 T is given at the right-hand side of the color-plot. The corresponding filing-factors are indicated.}
    \label{fig:Supplementary_Figure_1}
\end{figure}

In a Hall bar configuration, a clear distinction between longitudinal- ($R_{xx}$) and Hall-resistance ($R_{xy}$) can be made. As long as the Fermi energy is in between two Landau levels (LL) $R_{xx}=0$ while $R_{xy}=(h/ne^2)$. However, in our sample the situation is  more complex due to the absence of a graphene edge which directly couples to the contacts. As all four contacts are situated in a row, a separation between longitudinal- and Hall-resistance is impossible. Therefore, the 4-terminal resistance  $R_{14,23}$, shown in fig. \ref{fig:Supplementary_Figure_1}b, is more complicated to interpret and goes beyond the scope of this studies. The filling factors have been assigned based on a capacitance model.\\

The evolution of the filling factors with varying back-gate and magnetic field was determined using a non-local measurement as it revealed more pronounced features in the fan-plot. The back-gate voltage to density conversion ($n=\alpha V_{BG}$, where $\alpha$ is the lever-arm of the back-gate) was extracted from the evolution of the filling factors ($V_{BG}=B e v/(\alpha h)$, where $v$ is the filling factor of the LL and $h$ is the Planck constant) in the 4-terminal, non-local measurements shown in fig. \ref{fig:Supplementary_Figure_1}a. The evolution of the filling factors ($\nu=\pm$ 2, $\pm$ 6, $\pm$ 10...) shown in fig. \ref{fig:Supplementary_Figure_1}b are in good agreement with the expected sequence for single layer graphene, indicated with the red, dashed lines.\\

\begin{figure}[htbp]
    \centering
      \includegraphics[width=0.5\columnwidth]{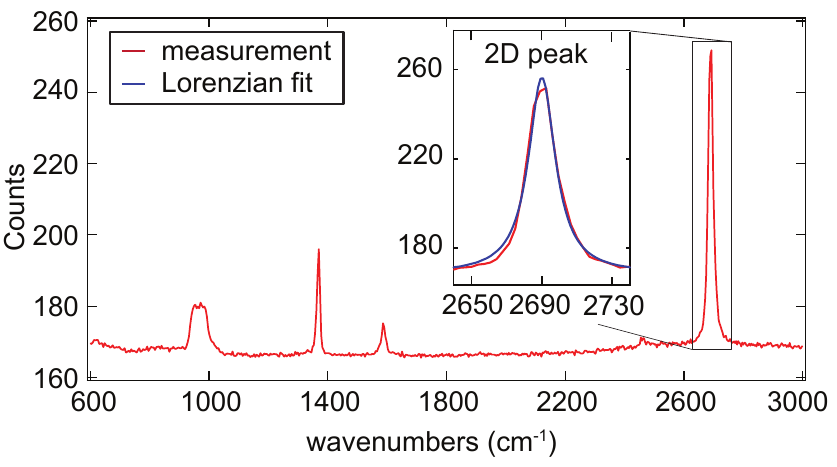}
    \caption{\textbf{Raman spectra of graphene encapsulated in hBN.} The exact number of graphene layers encapsulated in hBN was determined using Raman spectroscopy. The FWHM of the 2D-peak at roughly 2690 cm$^{-1}$ was fitted with a single Lorenzian (blue curve in the inset). The extracted FWHM of 18.3 cm$^{-1}$ fits best to the value expected for single layer graphene.}
    \label{fig:Supplementary_Raman}
\end{figure}

Fitting the full width at half maximum (FWHM) of the 2D-peak in the Raman spectra is an alternative way to determine the exact number of graphene layers present in the device, shown in fig. \ref{fig:Supplementary_Raman}. \cite{Hao09} We extracted a FWHM of  18.3 cm$^{-1}$ using a single Lorentzian to fit our  data which fits much better to the value of single layer graphene (FWHM=27.5$\pm$3.8 cm$^{-1}$) as compared to bilayer graphene (FWHM=51.7.5$\pm$1.7 cm$^{-1}$).\\

Having more knowledge about the evolution of the filling factors with back-gate and magnetic field, the various filling factors could be assigned to the peaks and valleys in fig. \ref{fig:Supplementary_Figure_1}b. A line-trace at $B$=5 T is given on the right-hand side of fig. \ref{fig:Supplementary_Figure_1}.

\subsection{Thermal cycling}
\begin{figure}[htbp]
    \centering
      \includegraphics[width=0.5\columnwidth]{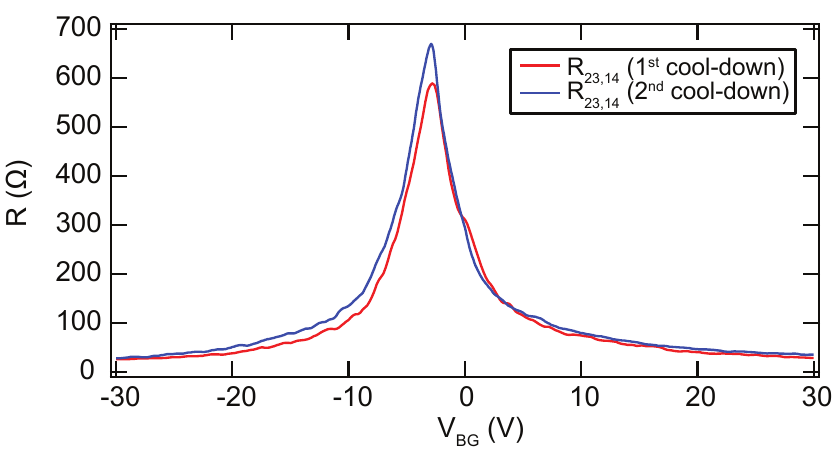}
    \caption{\textbf{Thermal cycling of point-contact devices} The devices did not change much during thermal cycling between the first and second cool-down as can be seen on the two measurements of the 4-terminal differential resistance. Thermal cycling was required to transfer the devices from the first measurement setup ($B\le$8 T) into a setup with high magnetic fields ($B\le$15 T).}
    \label{fig:Supplementary_Figure_2}
\end{figure}

The identical 4-terminal measurement was performed after the first- and second cool-down to obtain some knowledge about a possible degradation due to thermal cycling. As can be seen in fig. \ref{fig:Supplementary_Figure_2}, almost no change could be observed between the two measurements.
\newpage

\subsection{High field measurements}
\begin{figure}[htbp]
    \centering
      \includegraphics[width=1\columnwidth]{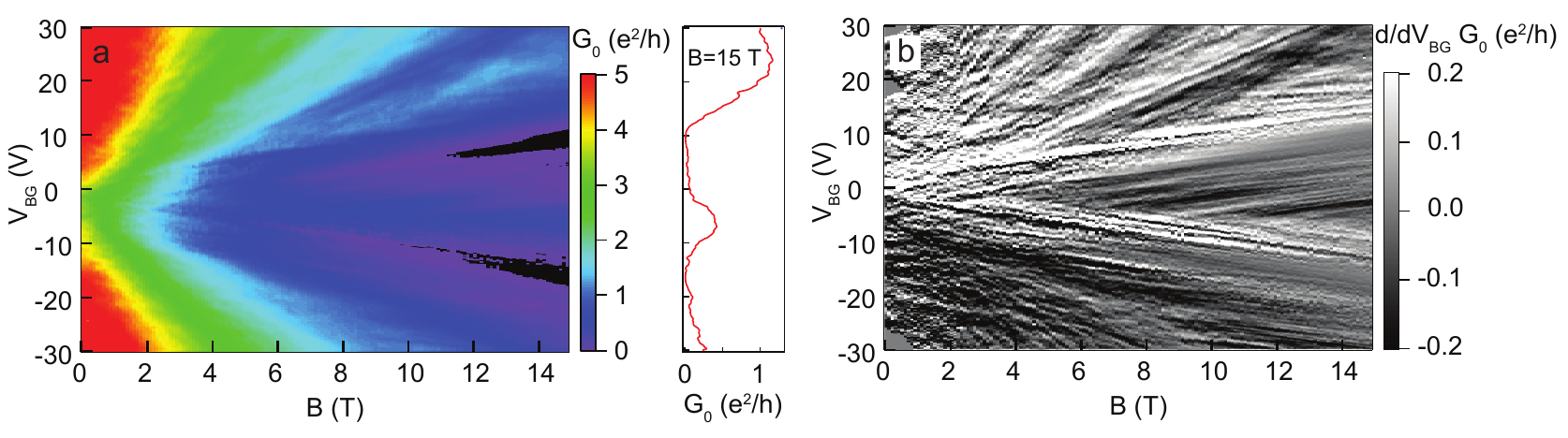}
     \caption{\textbf{2-terminal conductance  as a function of back-gate and magnetic field. a} The conductance between neighbouring PCs (distance of 2.2 $\mu$m) becomes insulating at high magnetic fields. A line-trace at $B$=15 T is given on the right-hand side of the color-plot. \textbf{b,} Numerical derivative ($d/dV_{BG} \; G_{0}(e^2/h)$) of the measurement given in (\textbf{a}).}
    \label{fig:Supplementary_Figure_3}
\end{figure}
High-field measurements of another sample with a distance of 2.2 $\mu$m in between the PCs is shown in fig. \ref{fig:Supplementary_Figure_3}.  It can be seen that the sample becomes insulating between neighbouring contacts at comparable fields as the sample shown in the main article.\\
 The black region in fig. \ref{fig:Supplementary_Figure_3}a shows the threshold for $G_0<0.025$ e$^2$/h ($R>$1 M$\Omega$).
 
\bibliographystyle{unsrt}

 \newpage

\end{document}